# HST OBSERVATIONS OF GRAVITATIONALLY LENSED FEATURES IN THE RICH CLUSTER AC114[1]

Ian Smail,[1] Warrick J. Couch,[2] Richard S. Ellis[3] & Ray M. Sharples[4]

1 Caltech 105-24, Pasadena, CA 91125.
2 School of Physics, University of New South Wales, PO Box 1, Kensington, NSW 2033, Australia.
3 Institute of Astronomy, Madingley Road, Cambridge CB3 0HA, UK.
4 Physics Department, University of Durham, South Road, Durham DH1 3LE, UK.

## Abstract

Deep Hubble Space Telescope images of superlative resolution obtained for the distant rich cluster AC114 ($z$=0.31) reveal a variety of gravitational lensing phenomena for which ground-based spectroscopy is available. We present a luminous arc which is clearly resolved by HST and appears to be a lensed $z$=0.64 sub-L$^*$ spiral galaxy with a detected rotation curve. Of greatest interest is a remarkably symmetrical pair of compact blue images separated by 10 arcsec and lying close to the cluster cD. We propose that these images arise from a single very faint background source gravitationally lensed by the cluster core. Deep ground-based spectroscopy confirms the lensing hypothesis and suggests the source is a compact star forming system at a redshift $z$=1.86. Taking advantage of the resolved structure around each image and their very blue colours, we have identified a candidate third image of the same source ∼50 arcsec away. The angular separation of the three images is much larger than previous multiply-imaged systems and indicates a deep gravitational potential in the cluster centre. *Resolved* multiply-imaged systems, readily recognised with HST, promise to provide unique constraints on the mass distribution in the cores of intermediate redshift clusters.

## 1 Introduction

Even in its pre-refurbishment state the Hubble Space Telescope (HST) was capable of high resolution imaging by virtue of the sharp core of its unique point spread function. We have started a deep imaging programme with HST designed to reveal the morphological nature of star forming galaxies in distant clusters (see Couch *et al.* 1994). The first target of this programme was the cluster AC114 ($z$=0.31) the data for which also reveals the remarkable potential of HST for identifying and studying gravitational lensing phenomena in intermediate redshift clusters.

AC114 is a well-studied cluster (Couch & Sharples 1987) with a high velocity dispersion ($\sigma_{cl} = 1649\pm220$ kms sec$^{-1}$), a compact core dominated by a cD and a ROSAT X-ray luminosity of $L_x \sim 4.0 \times 10^{44}$ ergs sec$^{-1}$. Its strong lensing power was first indicated in a survey for bright gravitationally lensed 'arcs' (Smail *et al.* 1991) where a candidate arc was identified at the largest cluster radius in the sample. The cluster was observed with the HST Wide Field Camera (WFC-I) in a series of exposures totalling 5.7 hours in a broad visual filter (F555W; $V_{555}$) and 6.1 hours in a broad near-infrared filter (F814W; $I_{814}$). The reduction of these frames is detailed in Couch *et al.* (1994). The final frames reveal structures on 0.1 arcsec scales with 1$\sigma$ surface brightness limits of $\mu_{555}$=27.7 and $\mu_{814} = 27.0$ magnitudes arcsec$^{-2}$. The HST images are complemented by a series of multi-colour UBVRI CCD frames and near-infrared $K$-band imaging of the cluster taken with the Danish 1.5m Telescope and the European Southern Observatory's 3.5m New Technology Telescope (NTT) at La Silla, as well as the 3.9m Anglo-Australian (AAT) and 4.0m Cerro Tololo Inter-American Observatory (CTIO) Telescopes.

In the next section we discuss the arcs discovered in this cluster by virtue of HST's high spatial resolution, including a new spectroscopically-confirmed arc. In §3 we present a unique gravitationally lensed system – a very wide separation multiply-imaged source. The largest separation between the candidate images is 50.6 arcsec – nearly an order of magnitude larger than the previous largest separation system. Finally, in §4 we give our conclusions and discuss the implications of this study for future observations of gravitational lensing phenomena with the post-refurbishment WFPC-II on-board HST. $H_o$=50 kms sec$^{-1}$ Mpc$^{-1}$ and $q_o$=0.5 are assumed throughout.

## 2 Gravitational Arcs

A shallow ground-based $I$-band image of AC114 is reproduced in Figure 1(a). Three arcs (A0-A2) were identified in the survey of Smail *et al.*(1991) and the higher spatial resolution of the WFC-I data confirms these and reveals a further 4 candidates (A3-A6, Figure 2). All these arcs are prominent on the combined NTT+CTIO deep $U$ image shown in Figure 1(b) which proves a remarkably effective way of identifying such blue sources because of their strong contrast against the intrinsically-red cluster galaxies. Generally speaking the arcs have optical and optical-infrared colours similar to those of giant arcs in other clusters (Smail *et al.*1993) and also similar to the colours of the faint field galaxy population (Table 1).

The unresolved nature of giant arcs in ground-based studies has been used to place limits on the concentration of the mass in the lensing clusters and the sizes of distant field galaxies (Wu & Hammer 1991). Although they are very faint, 5 of the 7 arcs appear unresolved even at the high spatial resolution of our WFC-I images. These observations push the resolution limit a factor of 5–10 smaller than these previous studies providing even more stringent joint limits on the concentration of the lensing cluster and the sizes of distant galaxies. We discuss the implications of this observation in the context of our detailed cluster mass model in Smail *et al.*(1994c).

The brightest and most striking arc is A0, which is spatially resolved by HST (Figure 2) with a bright elongated knot orthogonal to the direction of the cluster centre. The integrated colours indicate a $z \sim 0.6$–1.2 late-type spiral with the bluer knot possibly indicating a region of intense star formation. The axial ratio of the arc, lack of curvature and its large angular separation from the cluster centre imply that the original source is probably elliptical, orientated at a position angle close to the local shear direction (Kochanek 1990).

To check the lensing hypothesis for A0 we obtained a spectrum using the NTT. We used the EMMI spectrograph with a red-sensitive low dispersion grism and a Loral 2048$^2$ CCD. Exposures totalling 4.2 hrs were obtained over 2 nights (1993 September 14 and 15) in reasonably good conditions with seeing of $\simeq$1.0 arcsec FWHM. The coadded and sky-subtracted spectral image and extracted spectrum are shown in Figure 3(a) and 3(b) respectively. Two emission lines are visible in the spectrum, the more conspicuous is at $\lambda_{obs} = 6098$Å with the weaker line at $\lambda_{obs} = 8205$Å. Identification of these with [OII]$\lambda$3727 and [OIII]$\lambda$5007 respectively gives a source redshift of $z$=0.639. Of particular interest, and noticeable in Figure 3(a), is the significant tilt in the the emission features along the spatial direction. This velocity gradient, if it arises from disk rotation, confirms that the source must be preferentially orientated. At [OII]$\lambda$3727 the maximum shift across the image is 10Å (3 pixels) corresponding to a rotation amplitude of $\pm 245/\sin i$ kms sec$^{-1}$ at $z$=0.64. By assuming that the source is a late-type spiral and that the local Tully-Fisher relation (Pierce & Tully 1992) is valid at these redshifts, we can solve for the intrinsic inclination of the source, which appears in both the Tully-Fisher and lensing luminosity estimates. The intrinsic ellipticity we obtain is $\epsilon \sim 0.5$, compared to an observed value of $\epsilon \sim 0.7$, leading to an amplification factor of $\sim$ 1.5–2.5. The corresponding intrinsic luminosity is $L_V \sim 0.5L^*$ and the rotation curve spans $\sim$15–20 kpc at the source.

## 3 A Resolved Wide Separation Multiply-Imaged Pair

The most interesting candidate lensed system in AC114 is a pair of unusual images, about 12 arcsec north of the central cD (Figure 4(a)). Although separated by 10.0$\pm$0.2 arcsec, the images are highly symmetrical with a bright compact source (S1/S2) and a low surface brightness L-shaped extension (D1/D2). Even lower surface brightness features with similar morphology to D1/D2 but of larger extent are apparent just above the sky noise. Both objects have similar very blue optical and optical-infrared colours (Table 2), compatible with those of a high redshift star forming galaxy. In fact all the structures are blue, with D1/D2 being especially so. The extreme colours make these sources especially conspicuous on the deep $U$ image (Figure 1(b)). The two faint objects labelled E1 & E2 are much redder, with colours similar to those of cluster members. The high degree of morphological and photometric similarity leads us to suggest that S1+D1/S2+D2 represent two images of a single compact background source, possibly a star forming galaxy with a companion or jet.

Since HST *resolves* the image pair, we can determine if the *spatial* magnification ratio for the two images is consistent with the photometrically derived value. Aperture photometry from the F555W data yields magnitude differences of $\Delta_{12} = 0.16 \pm 0.01$ mag for S1/S2, and $\Delta_{12} = 0.14 \pm 0.03$ mag for D1/D2. The good agreement between these two estimates supports the lensing hypothesis and leads to a relative magnification ratio of 0.86$\pm$0.01. To measure the spatial magnification ratio we deconvolved the appropriately-processed F555W frames after sinc-interpolation onto a finer grid. The deconvolution was undertaken using the accelerated Lucy-Richardson algorithm (Lucy 1974) with an analytic point-spread function created with the TinyTim package (Krist 1992). The ratio of the separations S1–D1 and S2–D2 is 0.92±0.04. Given the difficulty of accurately measuring the transverse component of the magnification tensor for the two images we concede that while it further supports the lensing hypothesis, it is not conclusive.

To rigorously verify the lensing hypothesis for S1+D1/S2+D2 we need spectroscopic confirmation that they lie at the same redshift behind the cluster. S1 and S2 are just within reach of faint object spectrographs on 4m class telescopes and we have attempted to measure their redshifts during two observing seasons. In September 1992 we used the Low Dispersion Survey Spectrograph (LDSS-1) with a $1024^2$ thinned Tektronix CCD at the AAT. In a 6 hour integration in mediocre conditions we detected the continuum in S1 and S2 with a reasonable signal–to–noise ratio (S/N~ 5). However, the absence of strong emission lines in the spectra of such blue sources indicates that [OII]3727Å must be redshifted out of the LDSS-1 window (3700–7500 Å) leading to a redshift limit of $z \gtrsim 1$. Identification of a weak absorption feature at $\lambda_{\rm obs}$ = 4427Å with CIV$\lambda$1548Å provides a redshift of $z$=1.86. The strong similarity of the spectra indicated that they are likely to arise from the same source, in support of the lensing hypothesis.

In August 1993 the EMMI imaging spectrograph on the NTT was used to secure better S/N and higher dispersion spectra in the previously covered region and to explore the peripheral wavelength regions for emission lines. A variety of setups were used; these are summarised in Table 3. Again we detect no strong emission lines in the wavelength range 3000–8250 Å. However, a number of weak absorption features are visible in our blue arm grating (DIMD) spectra of *both* images (Figure 5), including confirmation of the feature at $\lambda_{\rm obs}$ = 4427Å seen in the AAT spectra. These weak absorption features are identifiable with well known absorption lines seen in the ultra–violet spectra of local extra-galactic HII regions (Rosa *et al.* 1984) and we show these identifications in Figure 5. The derived source redshift is $z$=1.86 and we believe the similarity of the spectra conclusively verify the lensing hypothesis.

To test for relative variability of S1+D1/S2+D2, we turn to the $V$ band ground-based data which was taken with the Danish 1.5m telescope as part of a search for distant supernovae (Nørgaard–Nielsen *et al.* 1989). The individual frames span an extended time baseline allowing us to measure variations in the relative magnitudes of S1+D1 and S2+D2. These would arise from intrinsic variability of the source coupled with the different path lengths for the two images through the lens. The ground-based data spans 37 months between 1985 and 1988 in 7 observations and shows no differences in the relative magnitudes above the measurement errors (0.10 mag). This is not surprising given that the data came from a search specifically targeted at detecting such variations and none were reported. The cluster has also been imaged with the Australia Telescope at 3cm and 6cm with no detections coincident with S1/S2 to a flux limit of $f \sim 65\mu$Jy in either map (Liang, priv. comm.). The lack of non-thermal emission supports our interpretation of the source as a strong star forming galaxy rather than an AGN.

The relative parity displayed by S1+D1 and S2+D2 indicate that the lensing mass is probably not located between them. The proximity of the system to the cluster core and cD, coupled with its extreme separation, implies that the cluster potential is likely to play a central role in creating this system, possibly in conjunction with the cluster cD. An independent estimate of the position of the centre of mass responsible for forming the various arcs in the cluster can be derived via Kochanek's method (Kochanek 1990): the resulting position is within 3 arcsec of the cD. To gain some feel for the distribution of mass in the cluster on larger scales we turn to the ROSAT X-ray observations of this cluster: an 18.6 PSPC exposure retrieved from the UK ROSAT Archive and our 15 ksec ROSAT HRI exposure. These two images have been combined and are overlayed on Figure 1(a) which shows the X-ray emission strongly peaked on the cluster cD. Furthermore, the orientation and ellipticity of the X-ray surface brightness distribution is similar to that shown on smaller scales by the cD's envelope. Weak lensing analysis of other rich clusters has shown that the cluster mass distribution is likely to be closely resemble the morphology of the X-ray gas and galaxy distributions (Smail *et al.* 1994b).

S1 and S2 appear to be separated tangentially relative to the centre of mass determined above and also have a small relative magnification implying no strong differential shear between the two images. Similar configurations arise for five-image models from elliptical mass distributions (Blandford & Kochanek 1987). It is thus of considerable importance to locate any further images of the source. In general, the other three images are fainter, with one located close to the lens centre and possibly strongly demagnified. In our case the important aid to identifying the other images is the unusual morphology and remarkable blue colours of S1/S2 which, together, form a distinctive signature.

We used the ground-based images (and the WFC-I data in the cluster core) to restrict the colours and the WFC-I F555W data to constrain the morphologies of suitable candidates. There are no unidentified blue objects within a circle centred on the cD and passing through S2. The most interesting objects picked up in this area are two faint extended structures parallel to S1/S2 and about 5 arcsec to the south east, we return to discuss these below. The lack of any bright images in the cluster core allows us to discard models which predict configurations involving three bright images ('cusp triples').

Within a radius of 1 arcmin (roughly the Einstein radius for a $z=1.86$ source given the cluster's measured velocity dispersion) there are only 4 suitable candidate images which satisfy our joint criteria (Table 2); these are labelled C1-C4 in Figures 1(b) and 4(b). The most promising of these are C3 and C4: C4 has better colour credentials, while C3 shows a remarkable morphological similarity to S1. Given the quoted photometric errors and the obvious structural resemblance of C3 to S1/S2 shown in Figure 4(b), we believe this is the more likely candidate. The photometric magnification of C3 is most accurately measured from the ground-based $U$ image. We determine a magnification ratio C3/S1+D1 = 0.47±0.01, consistent with the metric scale of the asymmetries seen in the WFC-I data (Figure 4(b)). The angular separation of C3 from the mid-point of S1/S2 is 48.7 arcsec.

## 4 Discussion and Conclusions

Multiply-imaged sources provide strong constraints on the shape and depth of the lensing potential. A multiply-imaged source with spatially extended features is especially useful for modelling since, in addition to image separation and relative magnification, the orientations, parities and components of the magnification tensor for the extended features must also be reproduced. In contrast for a giant arc the number of available constraints is usually much lower. This intrinsically limits the usefulness of surveys for giant arcs, although statistical analysis of large samples of arclets in a single cluster promise to provide new insights into the distribution of mass outside of the cluster core (Smail et al. 1994b; Bonnet et al. 1994).

With enough images and independent constraints on the shape, core size and position of the cluster mass the lensing potential may, ultimately, be so well-constrained as to make it possible to *invert* the lensing equations and assign approximate distances to other lensed sources in the field. In the case of AC114, this would, for example, include the arcs A1-A6 and the fainter pairs visible close to S1/S2, all of which are beyond reach of conventional spectrographs. Indeed, the pioneering analysis of gravitationally lensed features in the rich cluster A370 by Kneib and collaborators (Kneib et al. 1993) demonstrates the unique capabilities of these systems. They utilise a sophisticated model of the mass distribution in the cluster, motivated by modelling of the giant arc, to obtain the distances to other gravitationally lensed sources seen in the cluster core. For AC114 we have undertaken extensive modelling of the cluster mass distribution using all the available constraints, the results of which are presented in Smail et al. (1994c).

Even without such sophisticated models it is possible to derive some interesting results about the lensing cluster and the sources from our observations. For example, the fact that S1/S2 form an image pair rather than an arc means that the source must be extremely compact (see Fort & Miralda-Escudé 1993). This very steep light profile contrasts with those seen for the bulk of the faint field population (scale sizes of ∼0.5 arcsec) in high resolution ground– and space–based observations (Griffiths et al. 1992). However, we have another indication that high-$z$ sources are very compact from the unresolved widths of most of the arcs seen in this cluster. These two observations may be pointing towards a new class of high-$z$ galaxy or sub-galactic unit which, if numerous, would have profound effects on models of galaxy formation (c.f. Giraud 1992).

The mere existence of such a large separation multiply-imaged source indicates that the lens must be extremely massive and concentrated. To make a more quantitative statement we turn to the two spectroscopically confirmed lensed systems in the cluster, the arc A0 and the multiply-imaged source S1/S2/C3. Assuming the cluster can be modelled as a very simple singular isothermal sphere we can obtain two crude, but independent, estimates of it's mass. The first of these estimates comes from the angular separation of A0 from the lens centre, which for simplicity we take as the cD's position. The Einstein radius for a background source is:

$$b = \frac{4\pi\sigma_{cl}^2}{c^2}\frac{D_{ls}}{D_{os}} \sim 29'' \left(\frac{\sigma_{cl}^2}{1000\text{km/sec}}\right)^2 \frac{D_{ls}}{D_{os}}$$

where $D$ denotes the angular diameter distances between observer ($o$), lens ($l$) and source ($s$). The distance factor is $D_{ls}/D_{os} = 0.42$ for A0. Using the separation of the arc from the cD, ∼62 arcsec, and the amplification factor of 2.0±0.5 we then obtain an estimate of the effective velocity dispersion along the cluster's major axis of $\sigma_{cl} \sim 1600^{+150}_{-300}$ kms sec$^{-1}$, very close to the spectroscopically determined value. The second mass estimate uses the the separation of C3 from S1/S2, this should be roughly twice the Einstein radius of the cluster when corrected to the major axis. For a $z=1.86$ source $D_{ls}/D_{os} = 0.69$, the separation of C3-S1/S2 is ∼49 arcsec leading to an estimate of $\sigma \sim 1100$ kms sec$^{-1}$. Correcting this to the major axis using the cluster's ellipticity and position angle measured from the X-ray image gives $\sigma_{cl} \sim 1400$ kms sec$^{-1}$, again in reasonable agreement with the spectroscopic value. Thus both our estimates of the cluster mass support the spectroscopically determined value, indicating a very high mass for the cluster. This result is similar to those of other lensing studies of rich moderate redshift clusters (Kneib *et al.* 1993; Smail *et al.* 1994a). All of these studies indicate that, contrary to theoretical expectations, very massive clusters do exist at redshifts $z \gtrsim 0.2$.

The nature of the structure seen as D1/D2 is still in question. The likely high magnification of the source and its proposed redshift mean that we are observing S1+D1/S2+D2 on sub-kiloparsec scales. From the lack of non-thermal signatures in the spectra and the null radio detection it would appear that D1/D2 is not a jet. It is more likely that D1/D2 is a companion to the source S1/S2 which may be involved in a tidal interaction or merger – this in turn could trigger the apparently high star formation rate in S1/S2.

The possible applications of lensed pairs to estimate the surface density of very faint galaxies and to study their nature has been recently discussed by Fort & Miralda-Escudé (1993). A well defined sample of pairs promises to provide well constrained lensing models which can be inverted to estimate distances to any distorted background object viewed through the cluster. Deep imaging will provide such an ensemble of distorted objects which can be used to predict the redshift distribution of the background faint galaxy population (Smail *et al.* 1994a; Kneib *et al.* 1994). Evidently, more detailed information from deep ground-based and HST studies is required to fully exploit these remarkable systems. The principal conclusion of this paper is the promise of HST for future studies of this kind. Only with its superlative resolution would the strong symmetry of S1+D1/S2+D2 have been recognised.

## Acknowledgements

We thank staff at the Space Telescope European Coordinating Facility, particularly Hans-Martin Adorf, Bob Fosbury, Richard Hook and Gustaaf van Moorsel, for their expert assistance in processing this data. We also thank Professor Henning Jørgensen for providing follow-up imaging data, Alfonso Aragón-Salamanca for reducing the AAT $K$ images, Richard Bower for processing the ROSAT HRI image of AC114 and Alastair Edge for retrieving and processing the PSPC exposure from the UK ROSAT Archive. We acknowledge useful discussions with Roger Blandford, Richard Bower, Alastair Edge, Mike Fitchett, Francois Hammer, David Hogg, Chris Kochanek and Jordi Miralda-Escudé. This work was partially supported by SERC. IRS gratefully acknowledges a NATO Advanced Research Fellowship. WJC acknowledges financial support from the Australian Research Council.

**Figures**

**Figure 1:** (a) Ground-based $I$ exposure of AC114 ($z$=0.31) upon which is overlayed the smoothed ROSAT HRI+PSPC image. The dotted line encloses the field covered by Figure 1(b). (b) A deep $U$ exposure of the central regions of AC114 with objects referred to in the text marked. A comparison of Figures 1(a) and 1(b) illustrates the prominence of the lensed features in $U$ compared to the cluster members. At the redshift of the cluster 1 arcsec corresponds to 6 kpc. North is 16.4 degrees clockwise from the Y-axis in these and all subsequent figures.

**Figure 2:** The candidate arcs identified from the WFC-I imaging. The figures show unrestored F555W images of the candidate arcs, except for A0 which is a combined/restored ground-based $V$ and HST F555W image. For the arc A0 the knot (K) is marked. The radius vector to the cluster centre is orthogonal to the elongation of the knot. The major tickmarks on all the figures denote 1 arcsec. Two of the images have strips with no data across the boundaries of the WFC-I CCD chips.

**Figure 3:** (a) Sky subtracted spectral image of A0 around redshifted [OII]$\lambda$3727. Note the tilt to the emission line. Also shown is the observed wavelength of the emission line along the arc to illustrate the systematic variation with spatial position. The maximum rest-frame velocity shift is 245 kms sec$^{-1}$. The typical error in the wavelength measurement is 2Å. (b) Extracted spectrum of A0 with features providing the redshift identification of $z$=0.639 marked. The upper spectrum has been smoothed to the effective resolution of the observation and both spectra have had bad sky subtraction features cleaned.

**Figure 4:** (a) The bright image pair of the proposed multiply-imaged system as seen on the HST F555W frame after deconvolution using the Lucy–Richardson algorithm. The various structures referred to in the text are marked. (b) The candidate third images found on the basis of colours and morphology; the strongest candidate is C3. C3 is compared to a suitably demagnified copy of S1 highlighting the strong resemblance. In all figures the major tickmarks represent 1 arcsec.

**Figure 5:** The individual blue-arm long-slit spectra for S1+D1 and S2+D2 obtained with the NTT. Total exposure time is 34.0 ksec. Our identifications for the weak absorption lines seen in *both* spectra are shown; they yield a redshift for the source of $z$=1.86.